\documentclass[useAMS,usenatbib,aps,floatdix,nofootinbib]{mn2e}

\usepackage{graphics}
\usepackage{graphicx}
\usepackage{caption}
\usepackage{amssymb,amsmath}
\usepackage{amsmath}
\usepackage{psfrag}
\usepackage{graphicx}
\newcommand{\ve}[1]{\mathbf{q1}}
\usepackage{color}

\newcommand{\f}{\frac}

\newcommand{\be}{\begin{equation}}      
\newcommand{\ee}{\end{equation}}      
      
\newcommand{\bef}{\begin{figure}}      
\newcommand{\eef}{\end{figure}}      
\newcommand{\bea}{\begin{eqnarray}}    
\newcommand{\eea}{\end{eqnarray}}      

\newcommand{\av}[1]{\ensuremath{\left\langle q1 \right\rangle}}

\def\sposeq1{\hbox to 0pt{q1\hss}}
\def\ltapprox{\mathrel{\spose{\lower 3pt\hbox{$\mathchar"218$}}
\raise 2.0pt\hbox{$\mathchar"13C$}}}
\def\gtapprox{\mathrel{\spose{\lower 3pt\hbox{$\mathchar"218$}}
\raise 2.0pt\hbox{$\mathchar"13E$}}}
\def\inapprox{\mathrel{\spose{\lower 3pt\hbox{$\mathchar"218$}}
\raise 2.0pt\hbox{$\mathchar"232$}}}

\newcommand{\tve}[1]{\tilde{\boldsymbol{q1}}}

\def\bse{\begin{subequations}}
\def\ese{\end{subequations}}

\def\lsim{\raise 0.4ex\hbox{$<$}\kern -0.8em\lower 0.62ex\hbox{$\sim$}} 
\def\gsim{\raise 0.4ex\hbox{$>$}\kern -0.7em\lower 0.62ex\hbox{$\sim$}}

\def\f0N{f_0^{(N)}}
\def\bec{\begin{center}}
\def\eec{\end{center}}

\title[Triaxality from cold spherical systems] {On the generation of
  triaxiality in the collapse of cold spherical self-gravitating
  systems} \author[F. Sylos Labini, D. Benhaiem and M. Joyce]
      {Francesco Sylos Labini${^{1,2}}$, David Benhaiem${^{1,2}}$ and
        Michael Joyce${^{3,4}}$ \\ $^{1}$Centro Studi e Ricerche
        Enrico Fermi, Via Panisperna 89 A, Compendio del Viminale,
        00184 Rome, Italy \\ $^{2}$Istituto dei Sistemi Complessi
        Consiglio Nazionale delle Ricerche, Via dei Taurini 19, 00185
        Rome, Italy\\ $^{3}$UPMC Univ Paris 06, UMR 7585, LPNHE,
        F-75005, Paris, France\\ $^{4}$CNRS IN2P3, UMR 7585, LPNHE,
        F-75005, Paris, France}
\begin{document}

\date{\today}

\maketitle

\begin{abstract}
{Initially cold and spherically symmetric self-gravitating systems
  may give rise to a virial equilibrium state which is far from
  spherically symmetric, and typically triaxial. We focus here on how
  the degree of symmetry breaking in the final state depends on the
  initial density profile.  We note that the most asymmetric
  structures result when, during the collapse phase, there is a strong
  injection of energy preferentially into the particles which are
  localized initially in the outer shells. These particles are still
  collapsing when the others, initially located in the inner part, are
  already re-expanding; the motion of particles in a time varying
  potential allow them to gain kinetic energy { --- in some cases
    enough to be ejected from the system}. We show that this mechanism
  of energy gain amplifies the initial small deviations from perfect
  spherical symmetry due to finite $N$ fluctuations. This
  amplification is more efficient when the initial density profile
  {depends on radius, because particles have a greater spread of fall
    times compared to a uniform density profile, for which very close
    to symmetric final states are obtained}.  These effects lead to a
  distinctive correlation of the orientation of the final structure
  with the distribution of ejected mass, and also with the initial
  (very small) angular fluctuations.}
\end{abstract}


\section{Introduction}

That self-gravitating systems initially in highly spherically
symmetric configurations can relax to virial equilibria which break
this symmetry strongly has been known for several decades
\citep{Polyachenko_Shukhman_1981, merritt+aguilar_1985} and documented
since then by many numerical studies (see
e.g. \cite{aguilar+merritt_1990,theis+spurzem_1999,
  boily+athanassoula_2006,barnes_etal_2009,worrakitpoonpon_2014}).
This phenomenon, of formation, and argued to play a crucial role in
cosmological structure formation { (see e.g. \cite{huss_etal_1999,
    macmillan2006universal}) has come to be referred to as ``radial
  orbit instability" (ROI)}. This name has been adopted since such an
instability has been shown \citep{antonov_1961,
  Fridman+Polyachenko_etal_1984} to characterize spherically symmetric
stationary solutions of the collisionless Boltzmann equation with
purely radial orbits. Further it is plausible, as argued originally by
\cite{merritt+aguilar_1985}, that a similar mechanism is responsible
for the formation of triaxial structures observed starting from very
cold initial conditions, as in this case collapse tends to produce
strongly radial orbits. Different authors {(see references above)
  have discussed how the symmetry breaking develops during the
  evolution from both simple power law density profiles
  (e.g. \cite{boily+athanassoula_2006}) and from cosmological initial
  conditions (e.g. \cite{macmillan2006universal}.}

In this paper we consider how the degree of the final symmetry
breaking is related to the initial condition --- specifically to the
exponent of the initial density profile --- for the case of completely
cold initial conditions. Our focus on this aspect of the problem
allows us to elucidate the mechanism by which the symmetry breaking
actually occurs in the process of collapse from cold initial
conditions. More specifically, we show in detail how fluctuations
breaking spherical symmetry may be amplified by the very large energy
changes characteristic of the very violent relaxation from cold
initial conditions. This amplification is most effective when the
energy change a particle undergoes is both large and strongly
correlated with its initial radial position, leading to a maximal
effect from density profiles with intermediate exponents.  We
underline that the mechanism we identify { bringing about the
  amplification of the symmetry breaking operates far from equilibrium
  and has no apparent link to the ROI of equilibrium systems.}

\section{Numerical simulations}

For our study we have simulated numerically, using the N-body code
{\tt Gadget} \citep{springel_2005}, the evolution from initial
conditions in which $N$ particles are distributed randomly inside a
sphere following a radial density profile $\rho (r) \propto
r^{-\alpha}$, with $\alpha$ in the range $0\leq \alpha \leq 2.5$ {
  (the reasons for the choice of this range will be discussed below) }
The family of initial conditions are thus characterized by the two
parameters $\alpha$ and $N$.  We will focus here on the dependence on
$\alpha$ of the degree of symmetry breaking of the relaxed state. We
have also varied $N$ systematically (for each $\alpha$) in a range
from a few thousand to one hundred thousand particles, and {\it in
  this range} of $N$ our essential results and analysis are {weakly
  sensitive} to this parameter. We will report in future work a more
detailed investigation of the subtle (and numerically challenging)
issue of the asymptotic large $N$ dependence of spherical symmetry
breaking \citep{boily+athanassoula_2006, worrakitpoonpon_2014}.

All results presented here are for simulations in which energy was
conserved to within a tenth of a percent. { For simulations with
  $\alpha$ in the range $[0.25,2]$ this level of energy conservation
  has been attained using typical values of the essential numerical
  parameters in the GADGET code [$0.025$ for the $\eta$ parameter
    controlling the time-step, and a force accuracy of $\alpha_F=
    0.001$]. The cases in which $\alpha$ is outside this range are
  numerically more challenging because of singularities --- discussed
  further below --- both for $\alpha=0$ and $\alpha=3$ in the limit $N
  \rightarrow \infty$. For these cases we have subjected our results
  to additional {tests} of their robustness, checking their stability
  in particular to smaller time-steps (see also the discussion in
  \cite{joyce_etal_2009,Benhaiem_SylosLabini_2015}).  We have also
  studied carefully the effects of varying the force smoothing
  parameter (corresponding to its small scale at small distances), and
  {we} have found our results to be stable provided it is
  significantly smaller than the minimal characteristic size (see
  below) attained by the structure during its collapse For the
  simulations reported below the smoothing parameter is always in this
  latter range.}

\section{Results} 
\subsection{Collapse and Virialization} 
\begin{figure}
\vspace{1cm}
{
\par\centering \resizebox*{8cm}{7cm}{\includegraphics*{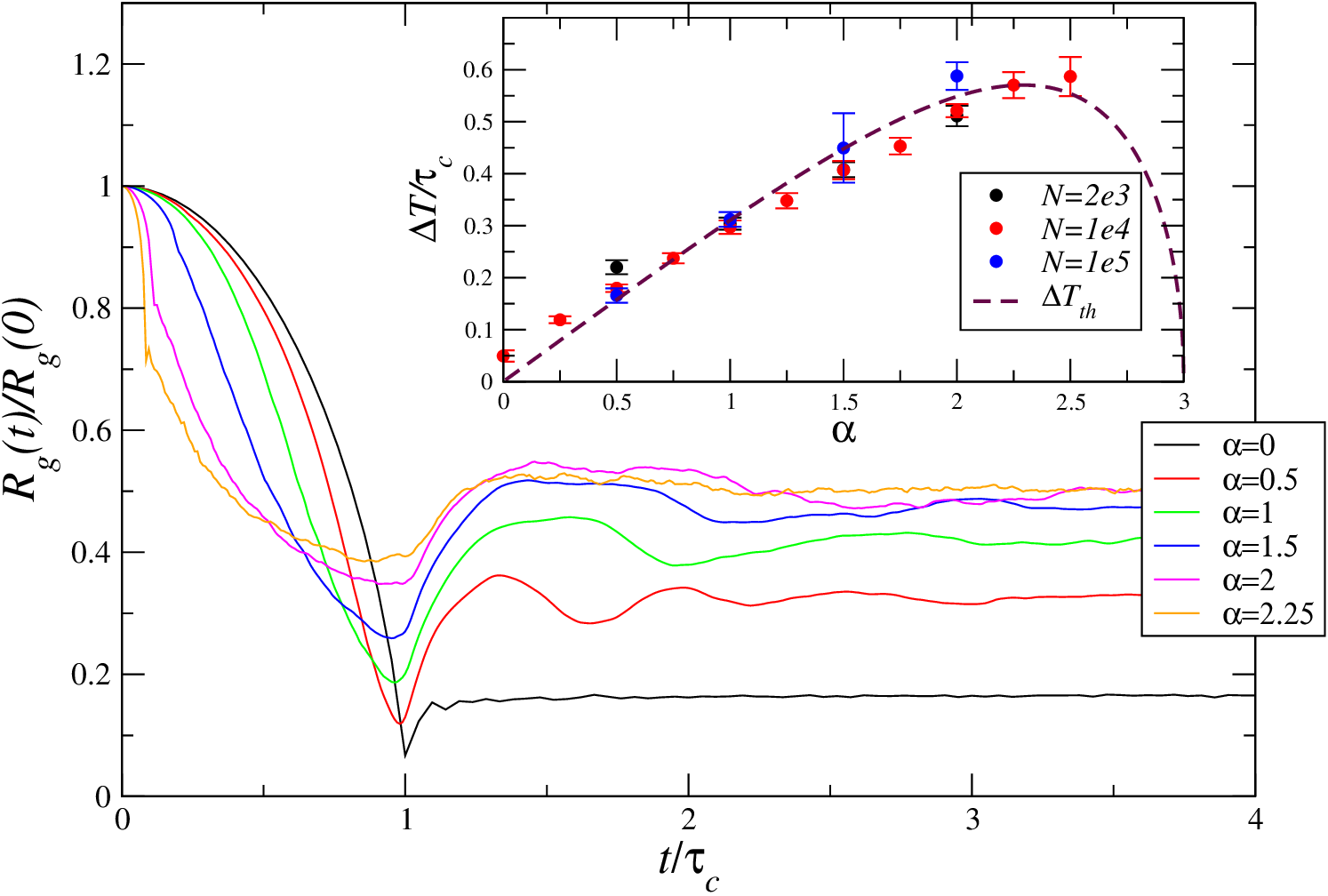}}
\par\centering
}
\caption{ 
{ Time evolution of the gravitational radius (as defined in
  Eq.\ref{Rg_def}) normalized to its initial value $R_g(0)$, for
  different $\alpha$ and with $N=10^4$. The inset shows the
  characteristic time of the spread in fall time $\Delta T$
  (normalized to $\tau_c$)
  in simulations with the indicated $N$; the dotted line corresponds
  to the analytical prediction given by Eq. (\ref{deltaT-alpha}).  } }
\label{Rgrav_powerlaw_1e4}
\end{figure}

 { Before turning to the central issue in this article --- the
   dependence of the asymmetry of the final virialized state on the
   exponent $\alpha$ --- we consider first how various spherically
   averaged indicators of the global evolution of the system vary with
   $\alpha$, and how these dependencies can be understood. The results
   in this section are in line with numerous previous studies of cold
   systems with initial conditions in the class we consider, or
   similar ones
   (e.g. \cite{vanalbada_1982,aarseth_etal_1988,joyce_etal_2009}).  We
   simply focus on the quantities and behaviors which will be most
   relevant for our analysis of the symmetry breaking in the next
   section.}
 
In line with previous studies of such initial conditions, we observe
that the system evolves through a strong collapse followed by a
re-expansion which very rapidly leads to a virial equilibrium in which
most of the initial mass is bound.  Fig.\ref{Rgrav_powerlaw_1e4}
shows, for different indicated values of $\alpha$, the temporal
evolution of the {\it gravitational radius} defined as \be
\label{Rg_def} 
R_g(t)  = \frac{G M_b(t)}{|W_b(t)|}
\ee
where $M_b(t)$ and $W_b(t)$ are respectively the mass which is bound
(i.e. particles with negative energy) and the potential energy of this
mass. The unit of time here is $\tau_c=\sqrt{ \frac { 3 \pi} {32 G
    \overline{\rho}_0(R_0)}}$, where $\overline{\rho}_0 (R_0)$ is the
{average mass density inside the radius $R_0$} of the initial
spherical configuration, of radius $R_0$.  It corresponds to the time
for a particle initially at the outer periphery (i.e. at $r=R_0$) to
fall to the center, in the continuum approximation (i.e. taking $N
\rightarrow \infty$ keeping the initial mass density profile fixed)
and without shell crossing.  The scale $R_g(t) $ is a measure of the
characteristic size of the system, and we observe in all cases, to a
first approximation, the same qualitative behavior --- a collapse to a
minimal size attained around $t=\tau_c$ followed by a re-expansion and
stabilization. {In all cases the bound particles form a virialized 
structure, with the virial ratio (which we do not display here) 
showing a very similar qualitative behaviour to that of $R_g(t)$
but with a final value close to $-1$ in all cases.} 
{There is, however, also a clear trend with $\alpha$: the smaller
  is $\alpha$, i.e. the closer to flat the density, the more violent
  is the collapse, with the system reaching a deeper minimum in a
  shorter time \footnote{{ We do not show data for $\alpha=2.5$ in
      Fig.\ref{Rgrav_powerlaw_1e4} because the potential energy $W_b$
      diverges for $\alpha \geq 2.5$ in the limit $N \rightarrow
      \infty$. As a measure of the characteristic size of the system
      we have used in this case the radius containing $90\%$ of the
      mass.} }. Further, we note that the larger is $\alpha$ the
  denser are the inner shells of the cloud and the sooner the collapse
  starts.}

 The variation of the characteristic time for the collapse and
re-expansion with $\alpha$ is quantified in the inset in the
figure. It shows, as a function of $\alpha$, the measured time $\Delta
T$ estimated as the difference between the two times at which
$R_g(t)=R_g^*$ defined as $R_g^*=(R_g^{asyn} + R_g^{min})/2$ where
$R_g^{asyn}$ is the estimated asymptotic value of $R_g$ at $t \gg
\tau_c$.  
{ The continuous curve, which has been extended to
  $\alpha=3$, is obtained from the following simple
  considerations.  We work in the approximation that departure from
  spherical symmetry, and also the effects of shell crossing, can be
  neglected. } 
For an initial mass density with radial profile $\rho(r) \propto
r^{-\alpha}$ for $r < R_0$, and zero for $r> R_0$, mass at an initial
radial distance $r_0$ from the center will then fall to the center in
a time $\tau_c (r_0)= \sqrt{ \frac {3\pi} {32 G
    \overline{\rho}_0(r_0)} } = \tau_c (r_0/R_0)^{\alpha/2}$, where
$\overline{\rho}_0(r_0)$ is the initial mass density of the sphere of
radius $r_0$.  The distribution $h(\tau)$ of these fall times to the
origin can then be calculated using $4\pi \rho(r_0) r_0^2 dr_0 = M
h(\tau) d\tau$ (where $M$ is the total mass). One finds
\be
\label{eq:htau} 
h(\tau) =  \frac{2(3-\alpha)}{\alpha \tau_c} \left(\frac{\tau}{\tau_c}\right)^{3(\frac{2}{\alpha}-1)}
\ee
for $\tau \leq \tau_c$ (and $h(\tau)=0$ otherwise). 
{The spread in the fall times can be characterized by the variance
  of $h(\tau)$, \be
\label{deltaT-alpha} 
\Delta T_{th} = 2\sqrt{ \langle \tau^2 \rangle - \langle \tau
  \rangle^2} = \frac{2 \alpha \tau_c}{6-\alpha}
\sqrt{\frac{(3-\alpha)}{3}} \;.  \ee This expression, plotted in the
inset of Fig. \ref{Rgrav_powerlaw_1e4}, reaches a maximum at $\alpha
\approx 2.3$, and goes to zero for both $\alpha=0$ and $\alpha=3$ .
For $\alpha=0$, all particles fall to the origin at $\tau_c$ (the
well-known singularity of the canonical ``spherical collapse model''),
while as $\alpha$ steepens towards $\alpha=3$ almost all of the mass
is {initially at} small radii with fall times which are very small
compared to $\tau_c$.  In the inset of Fig. \ref{Rgrav_powerlaw_1e4},
we see that Eq. (\ref{deltaT-alpha}) traces well the behavior of the
measured $\Delta T$ up to $\alpha \approx 2$.  Thus, up to this value,
the characteristic time of the variation of the total potential indeed
just reflects the spread in the particles' fall times.  Further for
larger $\alpha$ the behavior of $h(\tau)$ and $\Delta T_{th}$ indeed
reflect the qualitative change in behavior of the collapse we observe
in our simulations: while the collapse is completed only at $t \sim
\tau_c$ when the outermost mass falls, most of the mass falls at very
much shorter times. It is for this same reason that accurate numerical
integration becomes more costly as $\alpha$ increases and we report
results only up to $\alpha=2.5$.
}

\begin{figure}
\vspace{1cm}
{
\par\centering \resizebox*{7cm}{6cm}{\includegraphics*{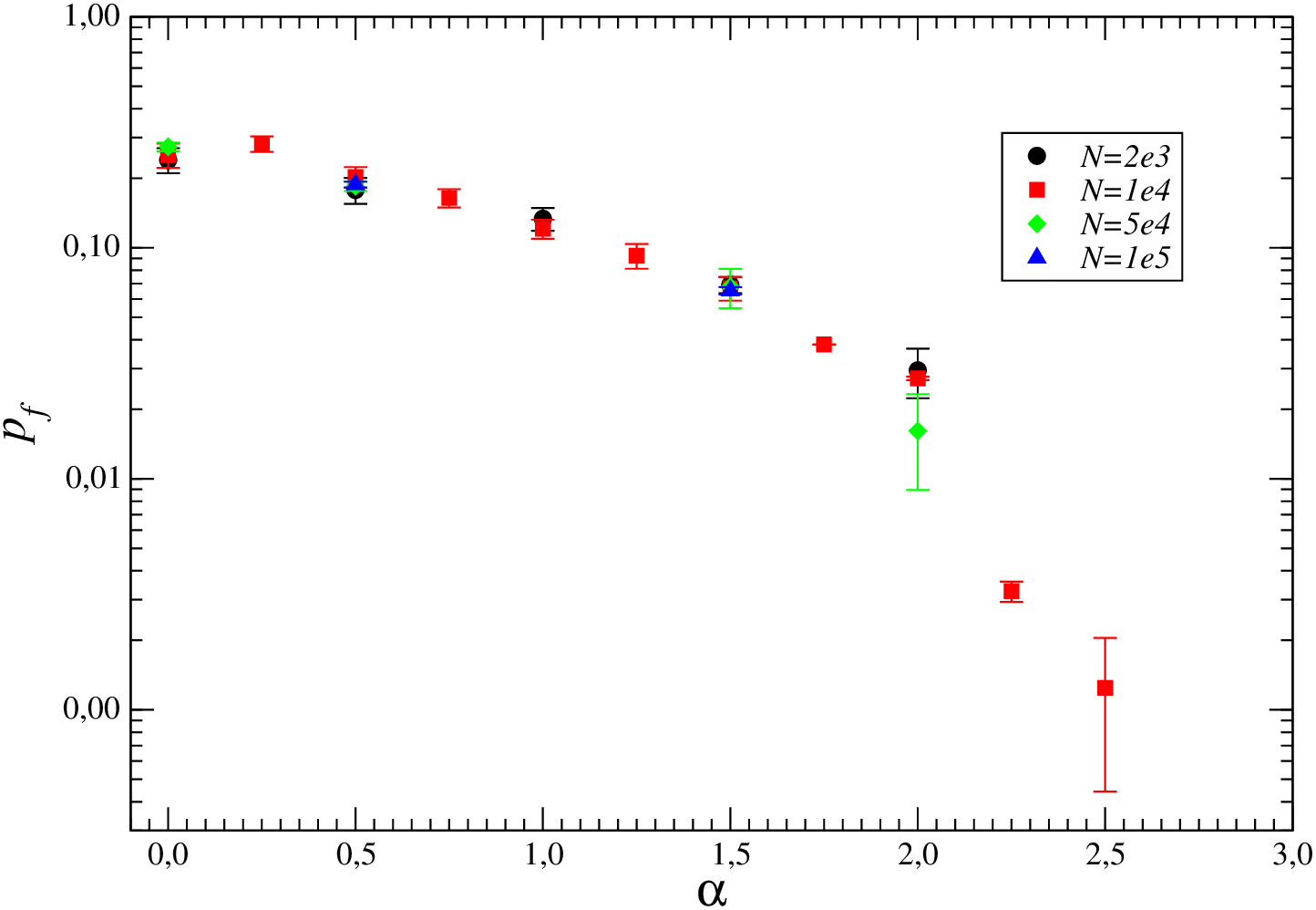}}
\par\centering
}
\caption{Fraction of ejected particles for different $\alpha$
{and different number of particles (see labels)}: average and
  standard deviation over $20$ realizations for $N=10^3$ and $N=10^4$,
  and $5$ realizations in the other two cases.}
\label{pf_alpha}
\end{figure}

One other macroscopic feature of relaxation from cold initial
conditions of this kind, which will be relevant in our discussion
below, is that they often lead to mass ejection i.e. some particles
gain enough energy so that their total energy is positive and they can
escape to infinity.  We show (see also \cite{syloslabini_2013}) in
Fig. \ref{pf_alpha} the fraction $p_f$ of the particles with positive
energy after relaxation (at $t \approx 5 \tau_c$), for different
$\alpha$ (and $N$).  The observed behavior as a function of $\alpha$
--- maximal ejection at $\alpha=0$, followed by a monotonic decrease
(approximately exponential) with $\alpha$ in a range until 
$\alpha \approx 2$, beyond which there is a much sharper drop --- is 
clearly related to the qualitative behavior of the fall times discussed
above. { This will be seen more explicitly in our analysis below. } 

\subsection{Symmetry breaking of relaxed state} 
\begin{figure}
\vspace{1cm}
{
\par\centering \resizebox*{8cm}{5cm}{\includegraphics*{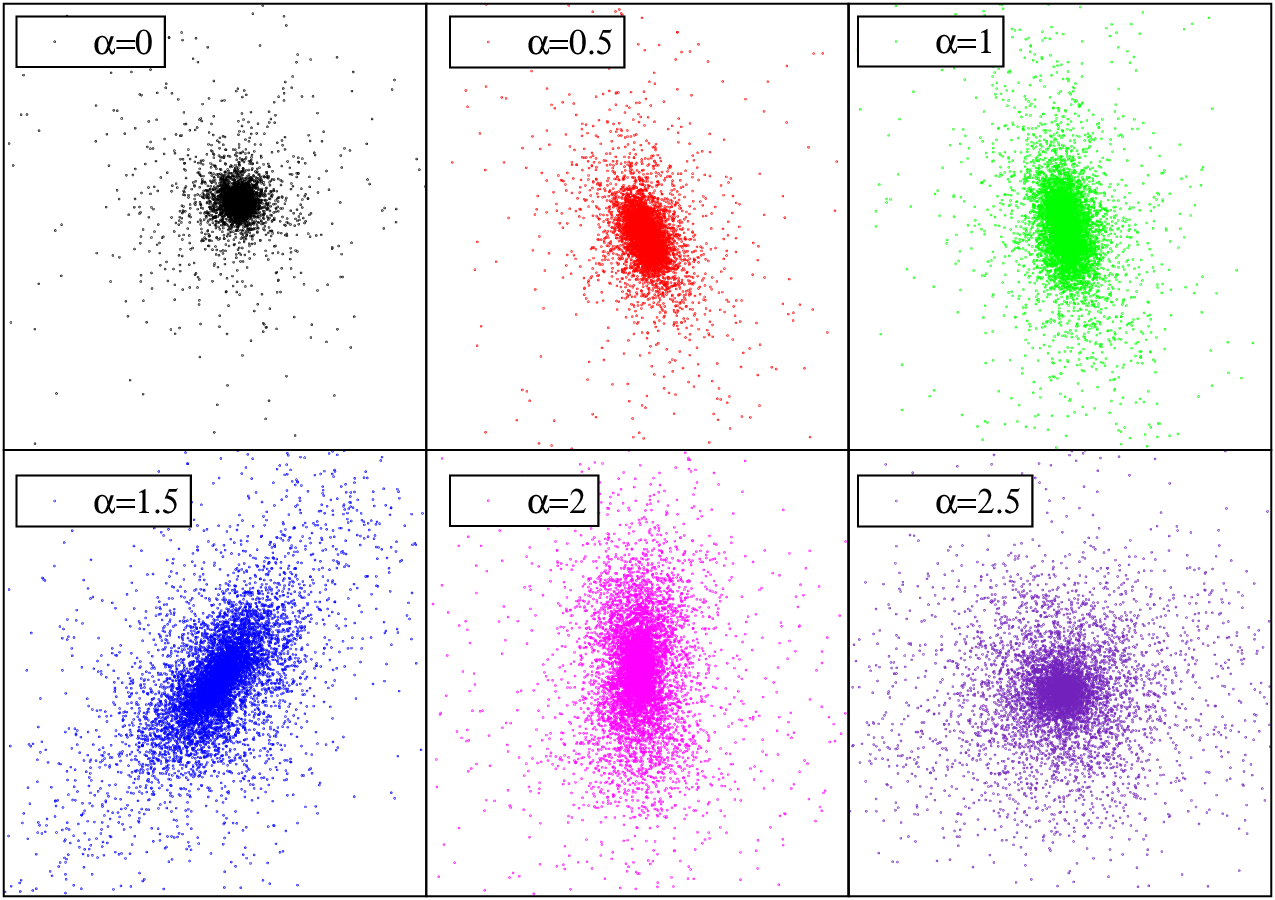}}
\par\centering
}
{
\par\centering \resizebox*{8cm}{5cm}{\includegraphics*{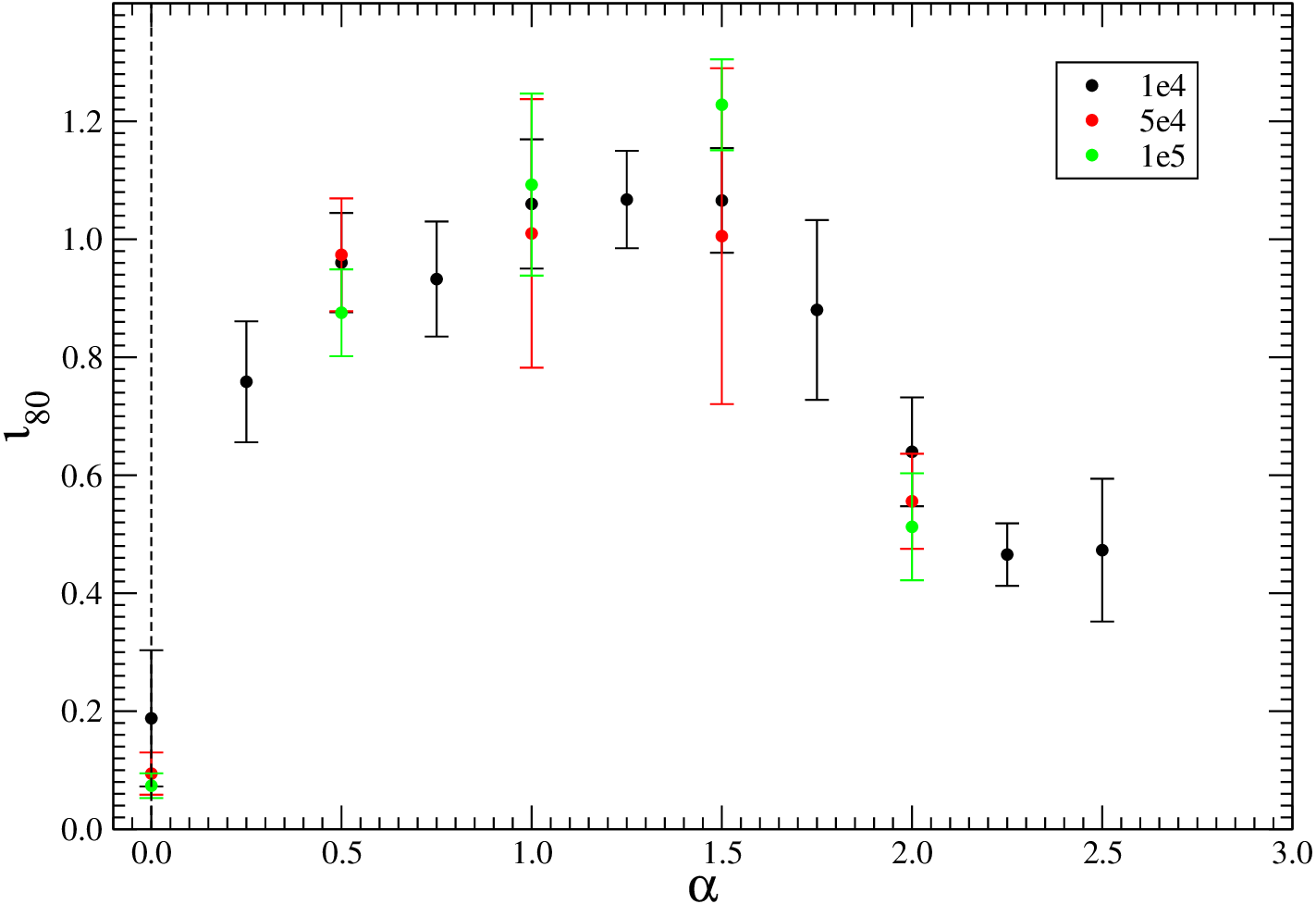}}
\par\centering
}
\caption{Upper plots: Projections of the virialized structure for a
  realization with $N=10^4$ for different $\alpha$ {at the final
    time $t \approx 5 \tau_c$}. Lower plot: flattening ratio
{(average and standard deviation)} $\iota_{80}$ as a function of
  $\alpha$, {estimated over 20 realizations for $N=10^3$ and
    $N=10^4$, and $5$ realizations in the other two cases. }}
\label{projection_alpha} 
\end{figure}

{ We focus now on the question of how the symmetry breaking in the
  final state depends on the exponent $\alpha$ characterizing the
density profile of the initial condition.}
 Shown in the upper
plots of Fig. \ref{projection_alpha} are a projection in a chosen
plane of the resulting virialized configurations for the various
indicated values of $\alpha$, for simulations with $N=10^4$. Visual
inspection suggests that the breaking of spherical symmetry is
apparently strongest in the intermediate values of $\alpha$, and
  weakest for the case $\alpha=0$.  This is confirmed by
Fig. \ref{projection_alpha} (lower panel) which shows, as a function
of $\alpha$, the parameter $\iota_{80}$ (the ``flattening ratio'') of
the relaxed state defined as
\be
\label{iota} 
\iota_P = \frac{\lambda_1}{\lambda_3} -1 
\ee 
where $\lambda_1$ and $\lambda_3$ are, respectively, the largest and
smallest of the three eigenvalues of the moment of inertia, and the
subscript indicates that the estimate is made on the $P$ \% of
particles which are most bound (we take here $P=80$ following common
practice in the literature).  We do not show here any information
about the intermediate eigenvalue $\lambda_2$, but analysis of it
shows that it is typically not close to the value of either of
$\lambda_1$ or $\lambda_3$, i.e. the relaxed structures are quite
triaxial. 
{ For each $\alpha$ the different points in the lower panel in
Fig. \ref{projection_alpha} correspond to the indicated values of $N$,
and the error bars to the standard deviation measured over the
indicated number of realizations in each case. The plot shows clearly
--- in agreement with previous studies of these initial conditions
(e.g. \cite{aarseth_etal_1988,joyce_etal_2009,barnes_etal_2009,worrakitpoonpon_2014})
--- that the final state in the case $\alpha=0$ is in fact very close
to spherically symmetric.  Further we note that above $\alpha \approx
1.5$ there is a clear trend towards progressively weaker symmetry
breaking, albeit with a lesser suppression than in the case
$\alpha=0$.  We have checked {and confirmed} that these trends of
$\iota_{80}$ with $\alpha$ are also observed with different values of
P in the calculation of $\iota_{P}$.}

{In the light of the discussion in the previous section, these
lots suggest a possible correlation between the observed asymmetry
and the qualitative behavior of the spread in the fall times.
Further the (relative) suppression at larger $\alpha$ suggests that
there may be some connection between the ejection of matter and the
degree of symmetry breaking.}

How are particles' fall times related to the final state, and in
particular its asymmetry?  In the relaxation process from such cold
initial conditions, particles energies change greatly as they move in
the time dependent mean field.  Indeed this is the essence of
``violent relaxation" as originally described by \cite{lyndenbell}. It
can be verified directly, by tracking the energy of individual
particles, that the energy change of any given particle occurs
essentially as it passes through the center of the
collapsing/re-expanding structure.  This is the case simply because
the mean field is most intense and most rapidly varying at this
time. As a consequence this energy change depends essentially on the
time window in which the particle passes through this central region.
The correlation between initial radial position and fall time ---
which is strong except in the limit $\alpha=0$ where all particles
have the same fall time (modulo finite $N$ fluctuations) --- might
thus be expected to lead to a correlation between the energy change of
a particle and its initial radial position.

\begin{figure}
\vspace{1cm}
{
\par\centering \resizebox*{8cm}{7cm}{\includegraphics*{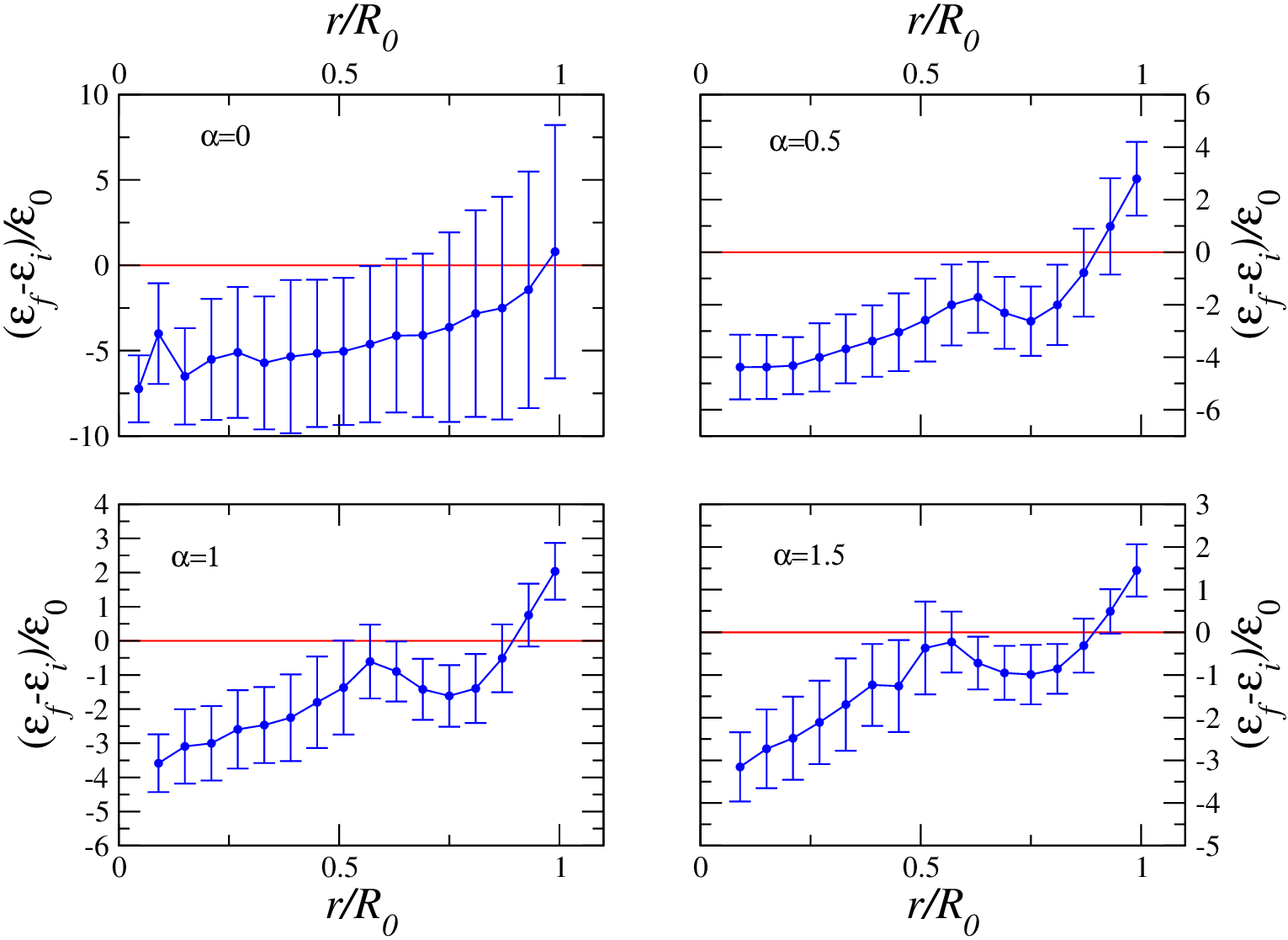}}
\par\centering
}
{
\par\centering \resizebox*{8cm}{3.5cm}{\includegraphics*{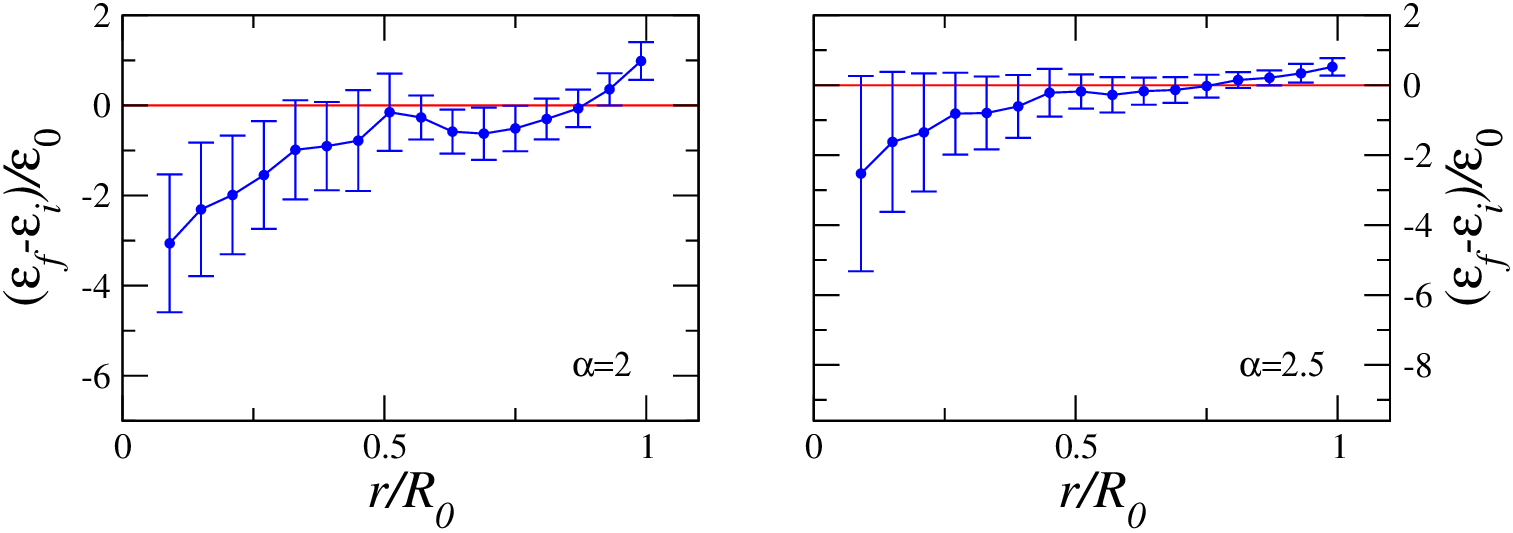}}
\par\centering
}
\caption{{ Change in particle energies {(average in bins)} in
    units of $\epsilon_0= G N^2 m/R_0$, plotted against their initial
    radial position $r_0$ (normalized to the initial maximum radius
    $R_0$), for the different indicated $\alpha$, and $N=10^4$. Note
    that the scale on the y-axis is different in the different
    plots.}} 
\label{DeltaE-r0} 
\end{figure}

To see whether this is indeed the case we plot in Fig.\ref{DeltaE-r0}
the total energy change, i.e. the difference between initial energy
and the energy measured at a time well after the collapse, when the
stationary state has been reached, { averaged in bins}, as a
function of their initial radial position, for simulations with
$N=10^4$.  
{ As anticipated we see that, except for the case $\alpha=0$,
there is indeed a very clearly identifiable correlation between energy
change and initial position: up to $\alpha=2$ a large positive energy
boost is {obtained} by a large fraction of the particles in the outer
shells, while at the larger values of $\alpha$ large energy decreases
are experienced by the particles in the inner shells. The explanation
for these features, and for the behavior in the case $\alpha=0$, is
closely related to our discussion of the previous section. As noted,
particles' energies change essentially as they pass through the center
of the structure, and what determines the energy change is the
temporal variation of the mean field they move in at this time.
}

The behavior of the mean-field potential at any point in the center of
the structure reflect approximately that of the total potential shown
in Fig.\ref{Rgrav_powerlaw_1e4}.  Particles which pass through the
center in the phase before the minimum is reached, at $t \sim \tau_c$,
will tend to lose energy because they climb out of a deeper potential
than they fall into, while the converse will be true for particles
which pass through the center as the system is re-expanding.  These
latter receive an energy boost, which has been noted to be at the
origin of the mass ejection in both the case of cold collapse
\citep{joyce_etal_2009,syloslabini_2012,syloslabini_2013} and in
merging structures \citep{carucci_etal_2014}.  
{Thus the trend we
  observe with $\alpha$ is clearly linked to that we discussed of the
  distribution of fall times $h(\tau)$: for small $\alpha$, a
  significant amount of the mass ``falls later" and acquires an energy
  boost, while at larger $\alpha$ most of the mass ``falls early" and
  loses energy. For $\alpha=0$, on the other hand, the correlation
  between the energy change of a particle and its initial position is
  very markedly weaker, because in this case the time of fall of a
  particle may become correlated with its initial radial position only
  through the finite $N$ fluctuations (which regulate the singularity
  of the collapse characteristic of this case).}

Let us consider now how the large energy injection into the outer
shells can lead to the very strong symmetry breaking of the relaxed
states observed in these cases. In the limit of exact spherical
symmetry, the dispersion of the energy change at any given initial
radius in Fig. \ref{DeltaE-r0} should vanish, and the observed finite
dispersion is a consequence of the spherical symmetry breaking.
Indeed what this dispersion implies is that the energy injection at
these large radii still depends sensibly on the direction of arrival,
and not just the initial radial position.

\begin{figure}
\vspace{1cm}
{
\par\centering \resizebox*{8cm}{7cm}{\includegraphics*{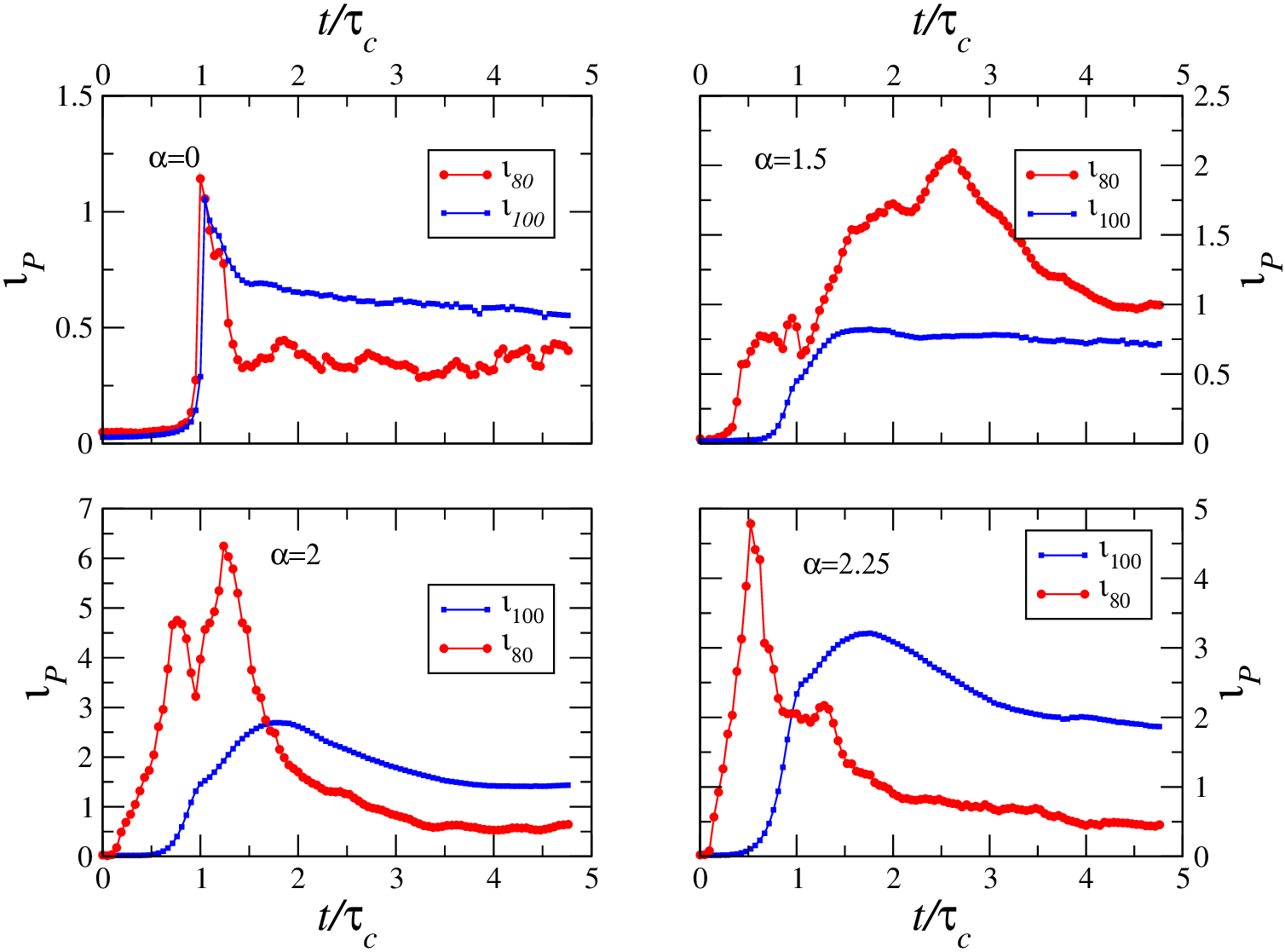}}
\par\centering
}
\caption{{ Behavior of $\iota_{80}(t)$ (circles) and $\iota_{100}$
    (squares, corresponding to all the bound particles) 
    as a function of time for different values of $\alpha\in [0,2.25]$ 
    , in simulations with $N=10^4$}.}
\label{M2a} 
\end{figure}  

{The initial system is not perfectly spherical due to Poisson
  fluctuations, and to a first approximation it can be described as an
  ellipsoid with $\iota(0) \sim N^{-1/2}$. Therefore particles
  experience, during the collapse, a force of which the tangential
  component depends on the angle with the ellipsoid initial major
  semi-axis. Consequently, as can be seen in Fig.\ref{M2a},
  $\iota_{80}(t)$ grows rapidly as the particles coming from the
  direction orthogonal to the initial major semi-axis collapse to the
  center in a shorter time than those along this axis: this is the
  \cite{Lin_Mestel_Shu_1965} instability whose effect is to amplify
  the initial small eccentricity. The particles initially in the
  outermost shells which arrive latest then travel through the rapidly
  decreasing potential created by the other mass which is already
  re-expanding, and gain energy.  Note that, while for $\alpha=0$ the
  amplification and subsequent decrease of the asymmetry to its final
  value occurs in a very short time around the collapse, for $\alpha
  >0$ the system shows a few oscillations before the complete
  relaxation.  This behavior reflects that of the gravitational radius
  in Fig.\ref{Rgrav_powerlaw_1e4}.  In addition, we note that for
  $\alpha=0$ the amplification of the initial $\iota(0)$ is much less
  marked compared to the cases with $\alpha>0$, in which the spread in
  fall times of the particles is much greater.  For $\alpha \ge 2$,
  most of mass now collapses at very short times and accordingly
  $\iota_{80}(t)$ shows a very fast growth for $t < \tau_c$ followed
  by a decrease when the external particles pass through the center.}
\begin{figure}
\vspace{1cm}
{
\par\centering \resizebox*{8cm}{7cm}{\includegraphics*{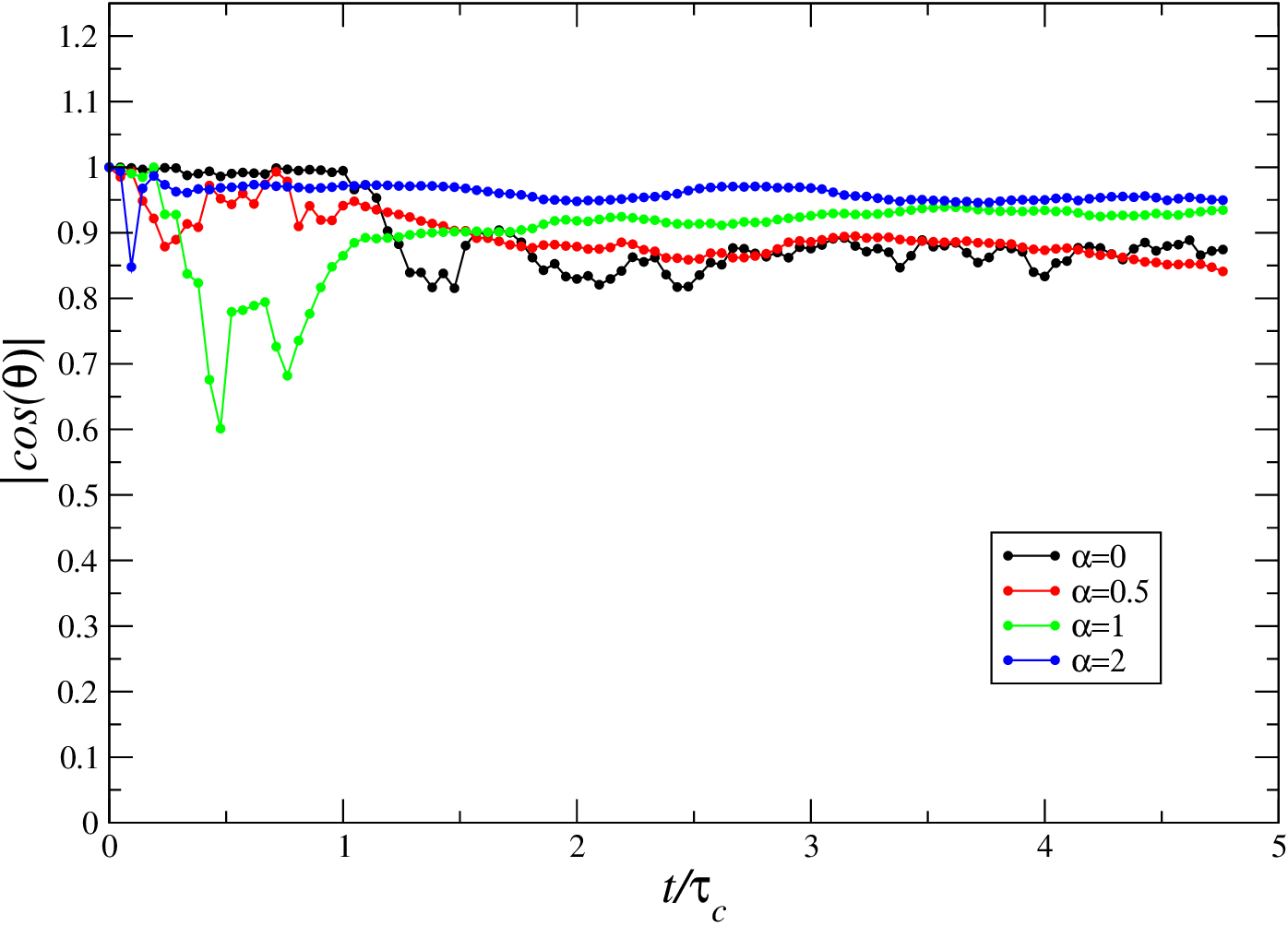}}
\par\centering
}
\caption{{
Behavior of absolute value of the cosine of
    the angle between the smallest eigenvalue (which corresponds to
    the major semi axis of the ellipsoid) at time $t=0$ and the
    smallest eigenvalue at time $t$, for different $\alpha$. It can
    be seen that it remains close to unity at all times, showing that
    the orientation of major semi-axis remains essentially unchanged
    throughout the evolution.}}
\label{M2c} 
\end{figure}

\begin{figure}
\vspace{1cm} { \par\centering
  \resizebox*{8cm}{7cm}{\includegraphics*{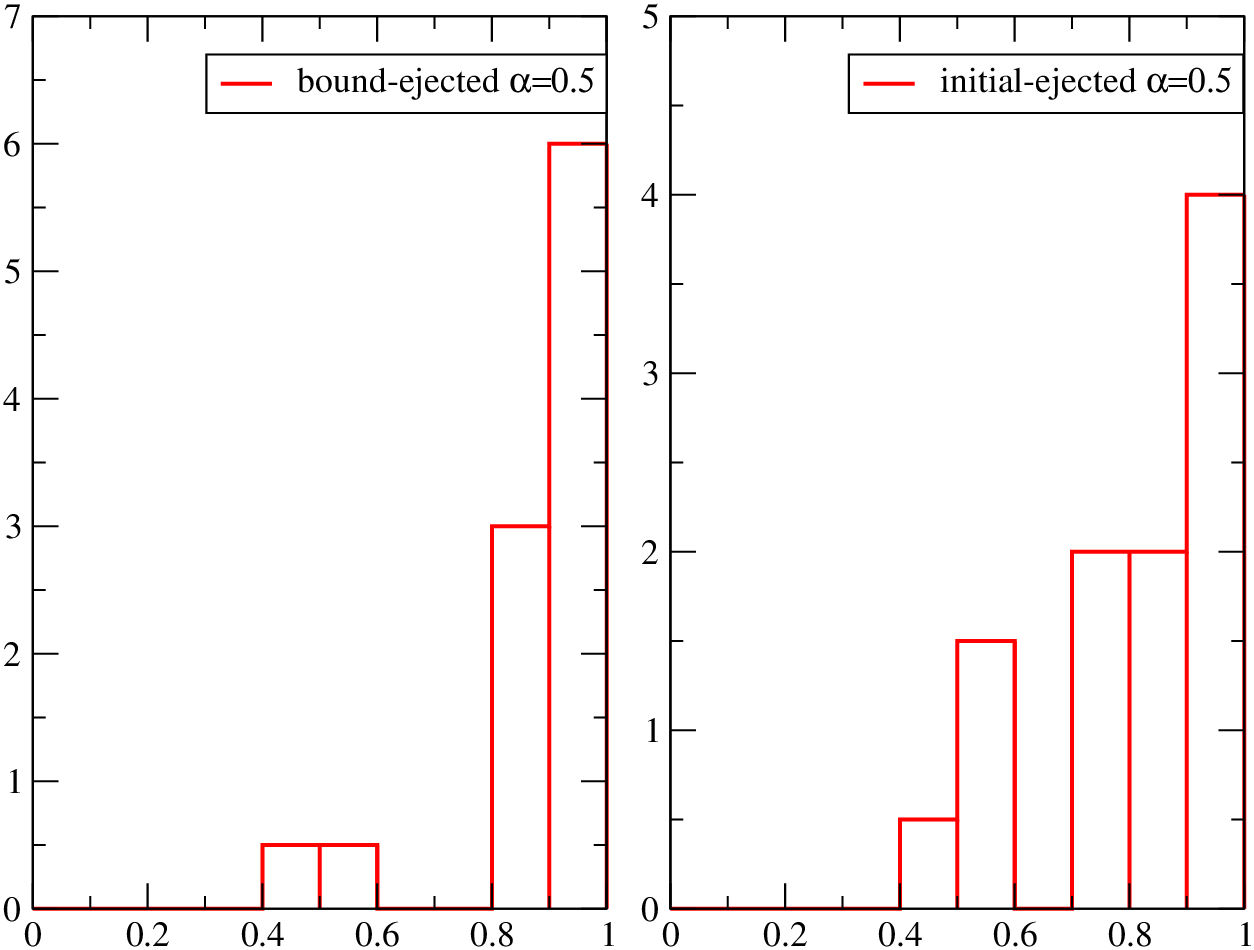}} 
\par\centering }
\caption{Histogram of the values measured, in 20 realizations
of the case $\alpha=0.5$ and $N=10^4$, of the cosine of the 
angle $\theta$ between the longest axes (determined from
inertia tensor)
of (i) the relaxed and ejected mass distributions (left panel),
and (ii) the initial and ejected mass distributions (right panel).} 
\label{costhetahisto_alpha_1e4}
\end{figure}

{ The hypothesis that the amplification of the initial triaxiality
  is due to finite $N$ effects via the instability described by
  \cite{Lin_Mestel_Shu_1965}, and the subsequent role of energy gain
  in amplifying the triaxiality, can be tested for
  straightforwardly. Firstly, we would expect that the major semi-axis
  at any time $t$ should be correlated to its value at $t=0$.
  Secondly, }when there is a significant mass ejection, we would
expect there to be a correlation between the angular distribution of
the ejected particles and the orientation of the final triaxial
structure, and more specifically between the preferred direction for
ejection and the elongated axis of the final structure.
{ As a measure of these spatial distributions we use simply the
  inertia tensor, determining its eigenvalues and eigenvectors.
  Fig.\ref{M2c} (upper panel) shows that the orientation of the major
  semi-axis indeed remains almost the same throughout the collapse and
  virialization}.
The left panel of Fig. \ref{costhetahisto_alpha_1e4} shows, for the
case $\alpha=0.5$, which has both a strongly triaxial final state and
significant mass ejection, a histogram of realizations of the modulus
of the cosine of the angle between the eigenvectors corresponding to
the smallest eigenvalues for the ejected mass and the $80\%$ most
bound mass.  There is again a very clear positive signal for the
correlation of the orientation of the axes.

The right panel of Fig. \ref{costhetahisto_alpha_1e4} shows the
correlation, measured in the same way, { for a number of
  realizations}, between the {\it initial} distribution and the
ejected mass.  We observe again a very clear correlation, and a
similar result is found considering the initial and final bound mass.
The reason for this strong correlation of the orientation of the
longest axes is simple: the moment of inertia of the initial mass
gives a measure of the small effective anisotropic due to the finite
$N$ fluctuations, and more specifically the axis of the smallest
eigenvalue is the axis along which the mass is ``stretched" furthest
away from the plane orthogonal to it passing through the center. It is
precisely along such a ``stretched" direction that one expects the
particles to arrive slightly later than the others --- just as the
mass along the longest axes of an ellipsoidal distribution --- and
thus to receive a slightly larger energy kick. This provides further
convincing evidence that the elongation of the final structure indeed
has its origin in the (relative) delay in particles' fall along these
directions.

{ This mechanism of symmetry breaking is more efficient for
  $\alpha$ in the range $[0.5,1.5]$ (see Fig. \ref{DeltaE-r0}). There
  is indeed greatest symmetry breaking in the final state for
  intermediate values of $\alpha$ where the energetically boosted
  particles are well localized in the outer radii, and where these
  same particles represent a significant fraction of the mass. The
  strong suppression of asymmetry in the case $\alpha=0$ is,
  conversely, due to the fact that, although there are many particles
  which pick up large energy boosts, they come from many different
  parts of the initial structure.  This leads to an effective
  averaging over the angular fluctuations and a much more spherical
  structure.}

\begin{figure}
\vspace{1cm} { \par\centering
  \resizebox*{8cm}{7cm}{\includegraphics*{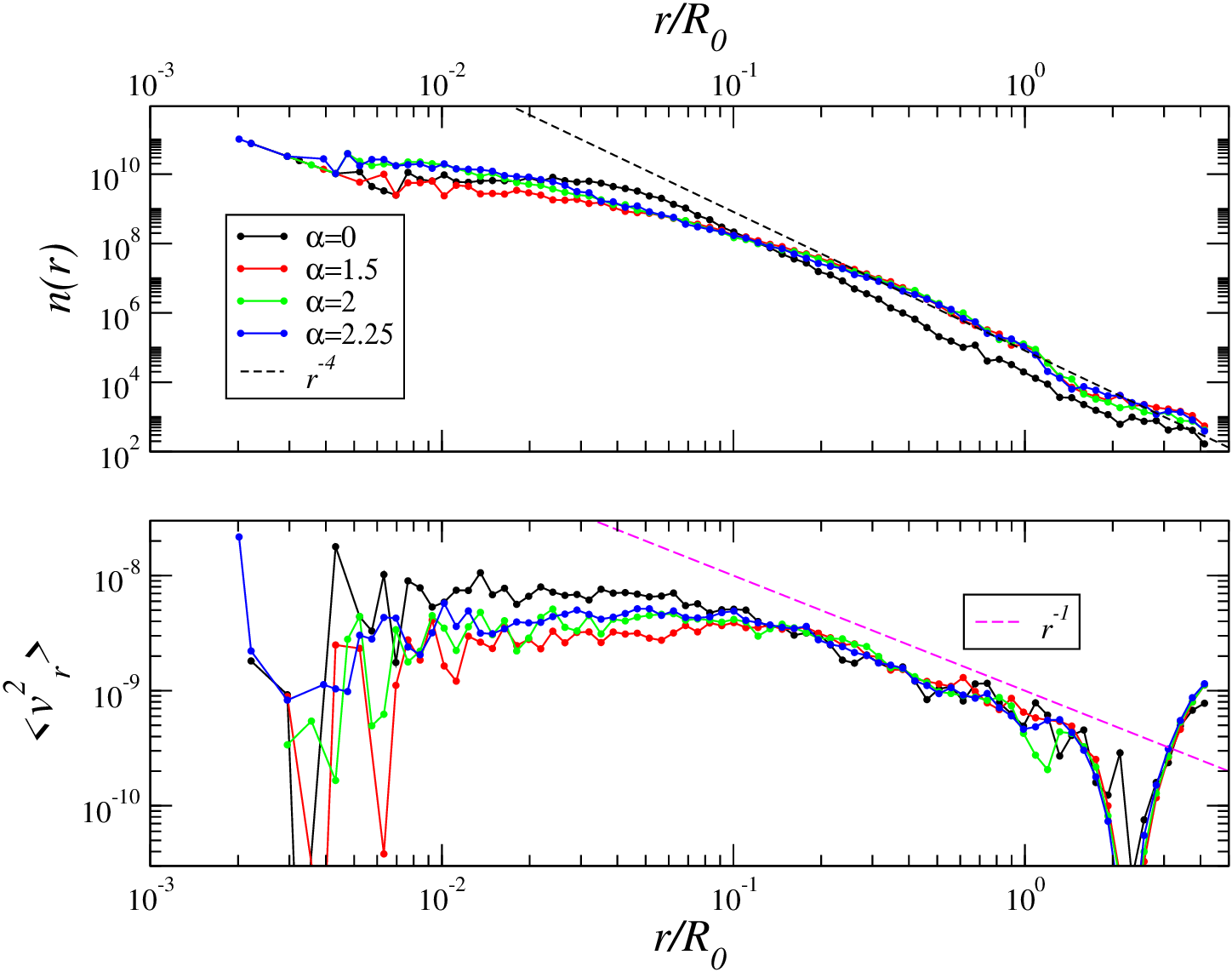}} 
\par\centering }
\caption{ Density (upper panel) and radial velocity (bottom panel)
  profiles for the final configuration at $t= 5 \tau_c$ and for
  different values of $\alpha$ { and with $N=10^4$.  The dashed
    line has a $r^{-4}$ behavior in the case of the density profile
    (upper panel) and a $r^{-1}$ behavior in the case of radial
    velocity profile.}}
      \label{densityprofile}
\end{figure}

{ We note that for $\alpha \ge 2$ there is a {markedly
    different behavior of $\iota_{100}$ and $\iota_{80}$.  This is a
    result of the fact, as seen in Fig.\ref{pf_alpha}, that there is
    no mass ejection in these cases: late-arriving particles do not
    gain enough kinetic energy to escape from the system.  In this
    case there are however a significant fraction of high energy (but
    bound) particles which reach large distances, giving rise to a
    configuration that is more asymmetric than that of the $80\%$ most
    bound particles.} Despite these differences the density and radial
  velocity profiles are very similar for all values of $\alpha$ (see
  Fig.\ref{densityprofile}), showing respectively a decay $n(r) \sim
  r^{-4}$ and $\langle v_r^2 \rangle \sim r^{-1}$ as noticed by
  \cite{syloslabini_2013}.  However the $\alpha=0$ case leads to a
  more compact configuration than $\alpha>0$ as a consequence of the
  larger spread of the energy distribution.
}

{We remark finally that a very similar dynamical process was observed
  by \cite{Benhaiem_SylosLabini_2015} for the case of initial
  conditions given by uniform ellipsoids. In addition, we note that
  \cite{theis+spurzem_1999} considered a Plummer density profile with
  a very small initial virial ratio, finding that a fast dynamical
  collapse generates at its end the maximal triaxiality of the system:
  the dynamical mechanism that generates such a triaxial structure is
  the same at work in the power-law density profile.}

\section{Discussion and conclusions} 

In summary, the mechanism for the generation of { the very strong
  spherical symmetry breaking observed for certain density profiles}
during violent relaxation from cold spherical initial conditions is
essentially the existence of a preferential axis for the large
``energy injection'' to particles in the outer parts of the initial
structure, which leads to an elongation of the final structure along
the same axis. This axis is itself defined by the finite $N$
fluctuations breaking spherical symmetry in the initial conditions,
corresponding to an axis along which the matter is on average further
from the origin. Particles initially at larger radii along this axis
fall through the collapsing region later than particles along the
other axes, and as a consequence pick up a larger energy injection.
After this time particle energies change negligibly, and the strong
variation with energy as a function of angle along the (predominantly
radial) orbits leads, after further phase mixing, to a virialized
structure with a spatial structure reflecting the energy injection. In
particular the { major semi-axis} of the final structure is
correlated both with the (very slightly) long axis of the initial
condition, and with that along which the energy kick obtained by
particles during violent relaxation is greatest. This mechanism has no
apparent relation to the instability of equilibrium systems with
radial orbits.  More specifically, the energy injection which is its
essential ingredient occurs in a very short time during violent
relaxation when the system is very far from equilibrium.  The system
never approaches close to an equilibrium configuration which is
spherically symmetric with purely radial orbits.

The analysis presented here is of completely cold initial conditions
only. { We can anticipate that both how the asymmetry of the final
  state, and the qualitative features of the collapse leading to it,
  will show the same behavior as a function of $\alpha$ for simple
  distributions of non-zero initial velocities,} provided the initial
virial ratio $b=2K/|W|$ associated is sufficiently small (see
e.g. \cite{syloslabini_2012}, which studies in particular the energy
injection and mass ejection as a function of $b$).  Given
\citep{Polyachenko_Shukhman_1981, merritt+aguilar_1985} that warm
(and, in particular, equilibrium) initial conditions are observed in
many cases to give rise to triaxial structures, by a mechanism which
clearly is intimately related to the ROI of equilibrium systems, we
are led to the conclusion that there are (at least) two distinct
mechanisms subsumed under what is usually called ROI. We note that
this conclusion is not only consistent with previous studies, but
gives an explanation of one of their striking (and puzzling) results,
namely that when the initial virial ratio $b$ is varied, there is a
critical value at which there is a qualitative change in behavior:
above this value symmetry breaking occurs only if the velocity
distribution is sufficiently radial, while below this value symmetry
breaking occurs irrespective of whether the velocity distribution is
isotropic or not (see e.g. Figure 4 of \cite{barnes_etal_2009}, which
locates the value at $b \approx 0.05 - 0.15$ depending on the
profile). 

While the dependence on the velocity anisotropic is a ``smoking gun"
for ``real" ROI operating for the warmer initial conditions, the
presence of a threshold below which the details of the velocity
distribution has no relevance is explained very naturally as the
dominance 
{ in this region of a distinct mechanism operating far
  from equilibrium, during the collapse, as we have described here.}
{A recent study} \citep{pakter_etal_2013} performs an analysis of the
  stability to elliptical perturbations of a uniform sphere
  (i.e. $\alpha=0$ here) with an isotropic velocity distribution and
  oscillating under its mean field, and finds that there is a critical
  value of the initial virial ratio below which there is such an
  instability.
{ Whether or not this is in fact the
  essential instability leading to deviation from spherical symmetry
  at early times during the collapses we have studied, the analysis of
  \cite{pakter_etal_2013} illustrates that, far from equilibrium,
  there are indeed such instabilities even when the velocity
  distribution is isotropic, and which are thus physically distinct
  from the ROI mechanism for equilibrium systems.} 

{We note finally that our analysis here has been for
isolated systems with initial power law initial conditions 
in a non-expanding universe.  A very similar phenomenology
of symmetry breaking starting from cosmological type
initial conditions in an expanding background has been
described  in \cite{macmillan2006universal}, and the
ROI in this case has been linked (see also \cite{huss_etal_1999})
to the generation of a ``universal''  NFW-type density profile
in this context. We do not observe the ROI in our study to be 
associated with such a final density profile: our profiles
are well characterised by a quite flat inner core
and power law  decay $\sim r^{-4}$.
To determine whether the specific mechanism of amplification
of symmetry breaking we have described here --- associated
with the energy injection to material initially in the outer 
shells --- is at play, or not, in the formation of triaxial
dark matter halos in cosmological models would require  
further extensive study,  incorporating a careful comparison 
between the dynamics of isolated structures and halos
in the cosmological setting.}
\bigskip 

Numerical simulations have been run on the Cineca PLX cluster (project
ISCRA QSS-SSG).  In addition, this work was granted access to the HPC
resources of The Institute for scientific Computing and Simulation
financed by Region \^Ile de France and the project Equip@Meso
(reference ANR-10-EQPX- 29-01) overseen by the French National
Research Agency (ANR) as part of the “Investissements d’Avenir”
program.

\end{document}